# DO NONEQUILIBRIUM PROCESSES HAVE FEATURES IN COMMON?

Leonid M. Martyushev
Institute of Industrial Ecology, Ekaterinburg, RUSSIA
mlm@ecko.uran.ru

The nature takes the easiest and most accessible paths and, hence, processes are accomplished very quickly in a minimum time. In 1662 P. Fermat used this principle to work out the refraction law. This was one of the first known attempts at successful deductive description of a physical phenomenon involving the variational principle. Presently researchers concerned with nonequilibrium processes have turned back to Fermat's idea in the form of the maximum entropy production principle (MEPP) (*1, 2*). In brief, a nonequilibrium system will most probably take the line of development when it maximizes the entropy production $\sigma$ at some assigned external constraints. The following relationship with Fermat's principle can be pointed out. It is known that the entropy production equals the product of the thermodynamic force $X$ by the flow $J$. Therefore, if, e.g., $X$ is fixed, the maximum entropy production leads to maximum $J$, i.e. selection of fastest processes. MEPP has proved to be good for understanding and description of diverse nonequilibrium processes in physics, chemistry and biology (*1, 2*). This brings up two questions: 1) Can this principle claim to be the basis of all nonequilibrium physics? 2) Is it possible to prove MEPP?

1. Recall the following points before we answer the first question. MEPP can be used to derive the whole apparatus of linear and, probably, nonlinear thermodynamics as it was shown by H. Ziegler. In the kinetic theory of gases MEPP can be used to derive the velocity distribution function for particles which satisfies the linearized Boltzmann equation and allows determining experimentally verified kinetic coefficients for gaseous, electron and phonon systems. Sufficiently fundamental studies can be found in the literature showing the relation of this principle to the Fokker-Planck equation, the method of nonequilibrium statistical operator and relaxation laws. These studies are reviewed in (*2*). Thus, the answer to the first question is "yes" today.

2. The second question is most interesting as follows from some recent studies (*2, 3*). Notice first that a principle like MEPP cannot be proved. Examples of its successful applications for description of observed phenomena just support this principle, while experimental results (if they appear) contradicting the principle will just point to the region of its actual applicability. The balance of the positive and negative experience will eventually lead to the consensus of opinion on the true versatility or a limited nature of MEPP. Other principles, such as laws of thermodynamics, Newton's law, etc., developed along similar lines. At the same time, there is always a temptation to relate the principle to other existing laws and, in this way, "prove" it. MEPP is traditionally related in the literature to the second law of thermodynamics (*2, 3*). However, investigators always have to use additional assumptions in their deductions and these assumptions prove to be less obvious than MEPP. Therefore, these proofs can hardly be viewed as fully successful. Still such efforts provide a deeper insight into the beginnings of MEPP and, for this reason, are always interesting and important. Let us adduce one more "proof" which, in our opinion, is most concise today. Assume that the second law of thermodynamics holds ($\sigma \geq 0$). Let $X = $ const $\geq 0$ and we have to demonstrate that the system selects $J$ (and, consequently, $\sigma$) as large as possible. Suppose several different flows are possible. All of them should be larger than zero because $\sigma \geq 0$ (the flows are directed towards the decrease of the thermodynamic force). However, since the reference system can be selected arbitrarily for the flows, we shall assume that the maximum flow is taken as the zero flow ($X$ should not change with this transformation because (a) their relationship with the flow is unknown in the case of the variational formulation and should be established (*2*) and (b) $X$ and $J$ have different dimensionalities). Then all the other flows are negative relative to the selected system leading to $\sigma < 0$. Since the second law of thermodynamics is a universal law of the nature and should not depend on such transformations, we prove that the maximum possible flow is realized at a given force and, hence, the entropy production is a maximum too.

Why is it MEPP that can be nominated for the general law describing nonequilibrium processes? Along with the above considerations, we shall point out two more factors which are important in P. Dirac's opinion for the "success" of a theory. They are simplicity and elegance. Both are rather obvious here. Many centuries ago researchers (for example, P. Fermat) turned to ideas similar to MEPP and these ideas appeared to them so obvious that were taken as axioms in development of a theory. The elegance of the theory follows from the fact that all equilibrium physics of many particles is based on maximization of the entropy (the method of potentials developed by Gibbs in thermodynamics and, for example, the well-known Jaynes' approach in statistical physics). Therefore the possibility that all nonequilibrium thermodynamics and statistical physics can be constructed on the basis of the entropy production (actually the time derivative of the entropy) maximization appears to be very intriguing.